\documentclass[11pt]{article}
\pdfoutput=1
\usepackage[a4paper]{geometry}
\usepackage{color}
\usepackage{float}
\usepackage[utf8]{inputenc}
\usepackage{a4wide}
\usepackage{amsmath,amssymb,amstext,amsthm,amssymb}
\usepackage{amsfonts}
\usepackage{framed}
\usepackage{graphicx}
\usepackage{bbold}
\usepackage{bbm}
\usepackage{slashed} 
\usepackage{tensor} 
\usepackage{braket} 
\usepackage{amstext}
\usepackage{array}
\usepackage{graphicx}
\usepackage{setspace}
\usepackage{hyperref}
\usepackage{overpic}
\usepackage{bbm}
\usepackage{cite}
\usepackage{lipsum}

\newcommand{\del}[0]{\partial}

\let\baraccent=\=
\renewcommand{\=}[1]{\stackrel{#1}{=}}

\newcommand{\eps}[0]{\varepsilon}

\begin{document}

\rightline{DESY-15-244}\vspace*{3ex}
\title{Towards Axion Monodromy Inflation with Warped KK-Modes}

\pagenumbering{gobble}
\begin{center}
{\huge \textbf{Towards Axion Monodromy Inflation with \\
\vspace{0.3cm}
Warped KK-Modes}}\\
\vspace{1.5cm}
{\large Arthur Hebecker$^1$, Jakob Moritz$^1$, Alexander Westphal$^2$ and Lukas T. Witkowski$^1$}\\
\vspace{0.5cm}
\textit{$^1$ Institute for Theoretical Physics, University of Heidelberg, \\
Philosophenweg 19, 69120 Heidelberg, Germany\\
\vspace{0.3cm}
$^2$ Deutsches Elektronen-Synchrotron, DESY, Notkestraße 85, 22607 Hamburg, Germany}\\
\vspace{1cm}
\textbf{Abstract}\\
\end{center}
\vspace{0.5cm}
We present a particularly simple model of axion monodromy: Our axion is the lowest-lying KK-mode of the RR-2-form-potential $C_2$ in the standard Klebanov-Strassler throat. One can think of this inflaton candidate as being defined by the integral of $C_2$ over the $S^2$ cycle of the throat. It obtains an exponentially small mass from the IR-region in which the $S^2$ shrinks to zero size both with respect to the Planck scale and the mass scale of local modes of the throat. Crucially, the $S^2$ cycle has to be shared between two throats, such that the second locus where the $S^2$ shrinks is also in a warped region. 
Well-known problems like the potentially dangerous back-reaction of brane/antibrane pairs and explicit supersymmetry breaking are not present in our scenario. However, the inflaton back-reaction starts to deform the geometry strongly once the field excursion approaches the Planck scale. We derive the system of differential equations required to treat this effect quantitatively. Numerical work is required to decide whether back-reaction makes the model suitable for realistic inflation. While we have to leave this crucial issue to future studies, we find it interesting that such a simple and explicit stringy monodromy model allows an originally sub-Planckian axion to go through many periods with full quantitative control before back-reaction becomes strong. Also, the mere existence of our ultra-light throat mode (with double exponentially suppressed mass) is noteworthy.

\vspace*{10ex}
\noindent December 14, 2015
\newpage
\pagenumbering{arabic}
\section{Introduction}
An important question in string cosmology is whether string theory compactifications allow for large-field inflation. On the one hand, many proposals for realizing inflation in string theory exist. At the same time, no-go theorems for large-field inflation have been put forward in various corners of the string theory landscape \cite{WGC,ConlonConstrains,NaturalInflationQuantumGravity,ConstraintsOnAxInf,TransplanckianAxions?,FencingInTheSwampland,PlanckinAxions,WindingOutOfTheSwamp,OnAxionicFieldRanges,WeakGravityStronglyConstrains,PaltiConstrains,WarpingWGC,ConstraintsKaloperKleban, 151203768}. By studying large-field inflation in string theory one may thus hope to learn about fundamental properties of string theory compactifications.

Furthermore, observation may force us to address these questions. For models of single-field slow-roll inflation, there is a direct link between the tensor-to-scalar ratio $r$ and the nature of inflation. To achieve $r \gtrsim 0.01$ the inflaton has to traverse a trans-Planckian field range during inflation, thus requiring inflation to be of large-field type \cite{Lyth:1996im}. The most recent observational constraint by BICEP2 and the Keck Array on the tensor-to-scalar ratio is $r\leq 0.07$ at 95 \% confidence \cite{TensorToScalar}, which is compatible with large-field inflation. Currently, considerable effort is being expended towards more precise measurements of $r$.

One challenge faced by models of large-field inflation is their sensitivity to an infinite tower of corrections to the inflaton potential. One way of controlling these corrections is to identify the inflaton with an axion-like field (henceforth axion), so that the shift symmetry of the axion protects the potential from dangerous corrections. A promising approach for realizing axion inflation in string theory is axion monodromy inflation \cite{MonodromyinCMB,GravitywavesLinearInflation}\footnote{See \cite{KaloperSorbo1,KaloperSorbo2} for an early, purely field theoretic version and \cite{KlebanUnwinding,Kleban2012,Kleban2014} for a more recent, closely related string-theoretic proposal.}. By introducing a monodromy the periodic field space of the axion is effectively unfolded, while the underlying periodicity of the theory continues to protect the inflaton potential from corrections. Further, an effective trans-Planckian field range for the inflaton can be achieved in theories involving more than one axion \cite{KNP, N-flation, Dante}. See \cite{Baumann:2014nda} for a review including advances until 2014. For more recent progress and further references see \cite{Westphal:2014ana}. 

A monodromy for axions can be induced by couplings to branes \cite{MonodromyinCMB,GravitywavesLinearInflation} (see \cite{MarchesanoD6} for very recent progress), but also due to background fluxes \cite{10114521} (recently established in the supergravity context under the name of $F$-term axion monodromy inflation \cite{F-term-inflation,14043542, 14043711}). All these approaches are not without their problems. For example, the original axion monodromy inflation constructions employ setups with both branes and anti-branes  \cite{MonodromyinCMB,GravitywavesLinearInflation}. As a result, in addition to the issue of back-reaction of the inflaton, the problem of brane-anti-brane back-reaction has to be addressed~\cite{Flauger:2009ab,BackreactionConlon} (see also \cite{PaltiWeigand}). Such models then require complicated warped throat geometries, which has hampered further quantitative studies of these constructions. Recent progress towards realizing such warped geometries has been made in \cite{BifidThroats}, where a $Z_2\times Z_3$-orbifold of the conifold is used. 

The situation is better in axion monodromy inflation models employing background fluxes, as the tools of flux compactifications can be used to examine these proposals in more detail. In \cite{14112032} it was shown that models of axion monodromy inflation in the complex structure moduli sector of Calabi-Yau 3- and 4-folds require a significant level of tuning to avoid excessive back-reaction and the destabilization of K\"ahler moduli. The required level of tuning can only be achieved in 4-folds which further complicates the model. These difficulties can be avoided if K\"ahler moduli are stabilized using non-geometric fluxes \cite{150301607, 150307634, 151001522}. However, it remains a challenge to implement a consistent hierarchy of scales in the resulting models.

Given the technical difficulties encountered in most constructions of axion monodromy inflation, it would be desirable to realize as minimal a model of axion monodromy inflation as possible. In such a simple construction one may hope that questions regarding the consistency and detailed phenomenology can be addressed explicitly and quantitatively. This is what we set out to do in this work. Here, we present a simple model of axion monodromy which is based on the standard Klebanov-Strassler-throat \cite{KlebanovStrassler} (i.e.~the deformed conifold) with shrinking $S^2$. Our axion is the RR-2-form $C_2$ wrapped on the homologically trivial $S^2$, similarly to some of the settings in \cite{F-term-inflation}. We do not need to include branes in our setup, the main point being that the axion acquires its monodromic potential from the homological triviality of the $S^2$ (in contrast to models where the potential is due to the tension of the NS5-brane). Thus we do not need to include anti-branes either and therefore evade the dangerous brane/antibrane back-reaction described in \cite{Flauger:2009ab,BackreactionConlon}. We note that our results might also be useful in the context of recently proposed Relaxion-models \cite{Relaxion1,Relaxion2,Relaxion3,Relaxion4,Relaxion5,Relaxion6,Relaxion7,151100132, Relaxion8,Relaxion9}.

We find that the mass of the lightest 4d-Kaluza-Klein mode is lighter than the next heavier mode by a \textit{relative} warp-factor which makes it an interesting candidate for single field inflation. Thus the inflaton potential is suppressed by warping \cite{UrangaWarpedThroats} without the need for an additional tuning. Since this is due to the $S^2$ ending in the infrared-region we need a second throat into which the $S^2$ can bend around in the UV such that its second end lies in an infrared region as well. Such a geometry has been constructed in \cite{IntriligatorVafaGeometry,AganagicVafaGeometry} which we very briefly review in Section \ref{Geometry}.

This paper is organized as follows: In Section \ref{10d-5d} we calculate the IR-localized 5d-mass-term, finding that $\Lambda\sim 1/R$ where $R$ is the typical radius of the KS-region. In Section \ref{5d-4d}, starting from the 5d-effective model we perform a Kaluza-Klein-reduction along the radial coordinate of the throat, thereby obtaining the effective 4d-theory with an infinite tower of KK-modes with the above mentioned mass-suppression of the lightest mode. In Section \ref{Checks} we compare the energy-densities of the inflaton with those stabilizing the throat, concluding that an explicit numerical back-reaction study is necessary to make statements about the stability of the KS-throat at large field excursion\footnote{Note that this is in contrast to a more optimistic claim of an earlier version of this paper.}. In Section \ref{section:UltraLightModeKS} the parametrization of the fully back-reacted inflaton mode in the KS-throat is given while the differential equations that need to be solved are listed in Appendix \ref{Appendix:EOM }. We draw our conclusions in Section \ref{Conclusion}.
\section{The Double Throat}
\label{Geometry}
\begin{figure}
\centering
\includegraphics[keepaspectratio, width=7cm]{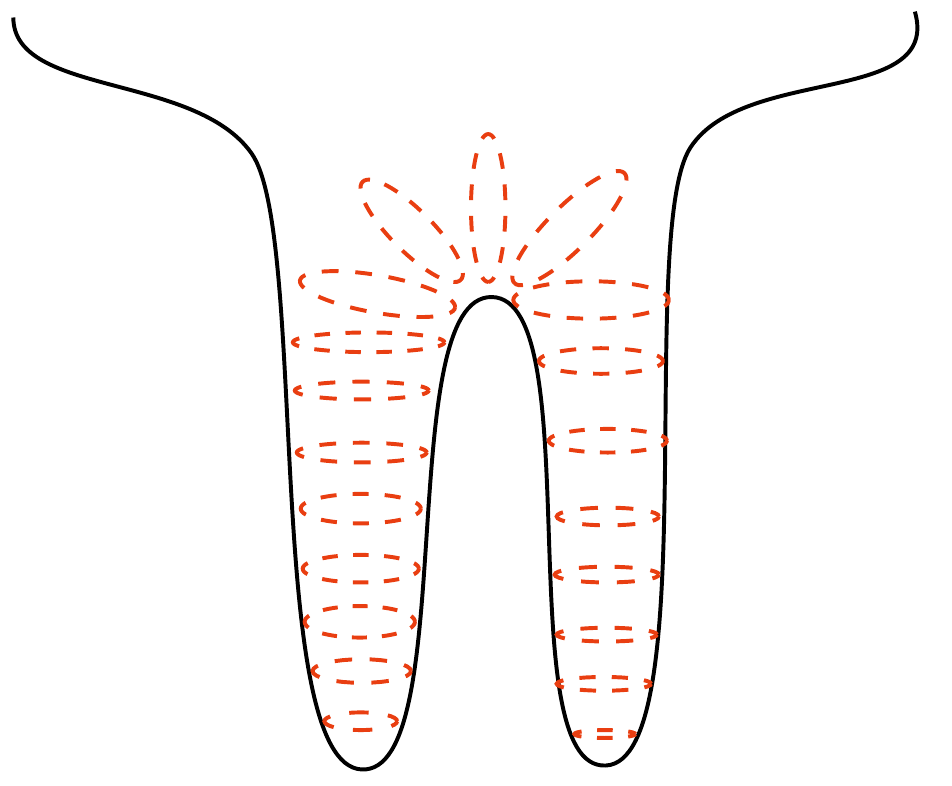}
\caption{The double throat: The dashed line indicates the family of $S^2$'s bending around into the second throat, shrinking to zero size at the tips.}
\label{fig:DoubleThroat}
\end{figure}

Let us briefly review the construction of the double throat (see Figure \ref{fig:DoubleThroat}) following the discussion in \cite{AganagicVafaGeometry}\footnote{This geometry can also be viewed as a $Z_2$-orbifold of the conifold \cite{BifidThroats}.}. The conifold can be described as the subset of $\mathbb{C}^4$ solving
\begin{equation}
\label{conifold}
uv=y^2+x^2\quad ,\quad
(u,v,x,y)\in \mathbb{C}^4\; .
\end{equation}
The conifold singularity sits at $x=y=u=v=0$. We can construct a two-conifold-setup by replacing $x$ with a polynomial $W'(x)$ in the conifold equation \eqref{conifold}. We take $W'$  to have two simple roots at $x\in\{a_1,a_2\}$:
\begin{equation}
uv=y^2+W'(x)^2\quad\text{where}\quad W'(x)=g(x-a_1)(x-a_2)\; .
\end{equation}
If $g=0$ this gives a curve of $A_1$-singularities parametrized by $x$. Blowing up the singularity gives a curve of $P^1$'s. Setting $g\neq 0$ there is still a family of $S^2$'s related in homology. After a geometric transition \cite{IntriligatorVafaGeometry} the system is deformed by means of a polynomial $f_1$ of degree one, to give two deformed conifolds with shrinking $S^2$:
\begin{equation}
uv=y^2+W'(x)^2+f_1(x)\; .
\end{equation}
This is precisely the geometry we will use.
\section{A Simple Geometric Setup and Reduction to 5d}
\label{10d-5d}
Consider the standard KS-throat with a blown up $S^3_{KS}$ but trivial $S^2$-cycle. Due to the homological triviality of the $S^2$ there is no harmonic 2-form and thus no massless axion $c= \int_{S^2}C_2$. Our axion will hence be the (massive) lightest KK-mode of $C_2$. \par
As a first approximation, let us take the geometry of the compact space to be simply $M_6=S^3_{KS} \times X_3$ where $X_3$ is a `cylinder' ($=S^2 \times \mathbb{R}$) of constant radius $R$, which is closed by one half of a three-sphere ($\equiv S^3_{1/2}$) in the IR. In the UV the $S_2$ bends around into a second throat such that it is closed in the IR on both sides as depicted in Figure \ref{fig:DoubleThroat}. This is crucial since we would otherwise generate a UV-mass-term.\par
\begin{figure}
\centering
 \begin{overpic}[width=0.40\textwidth,tics=10]{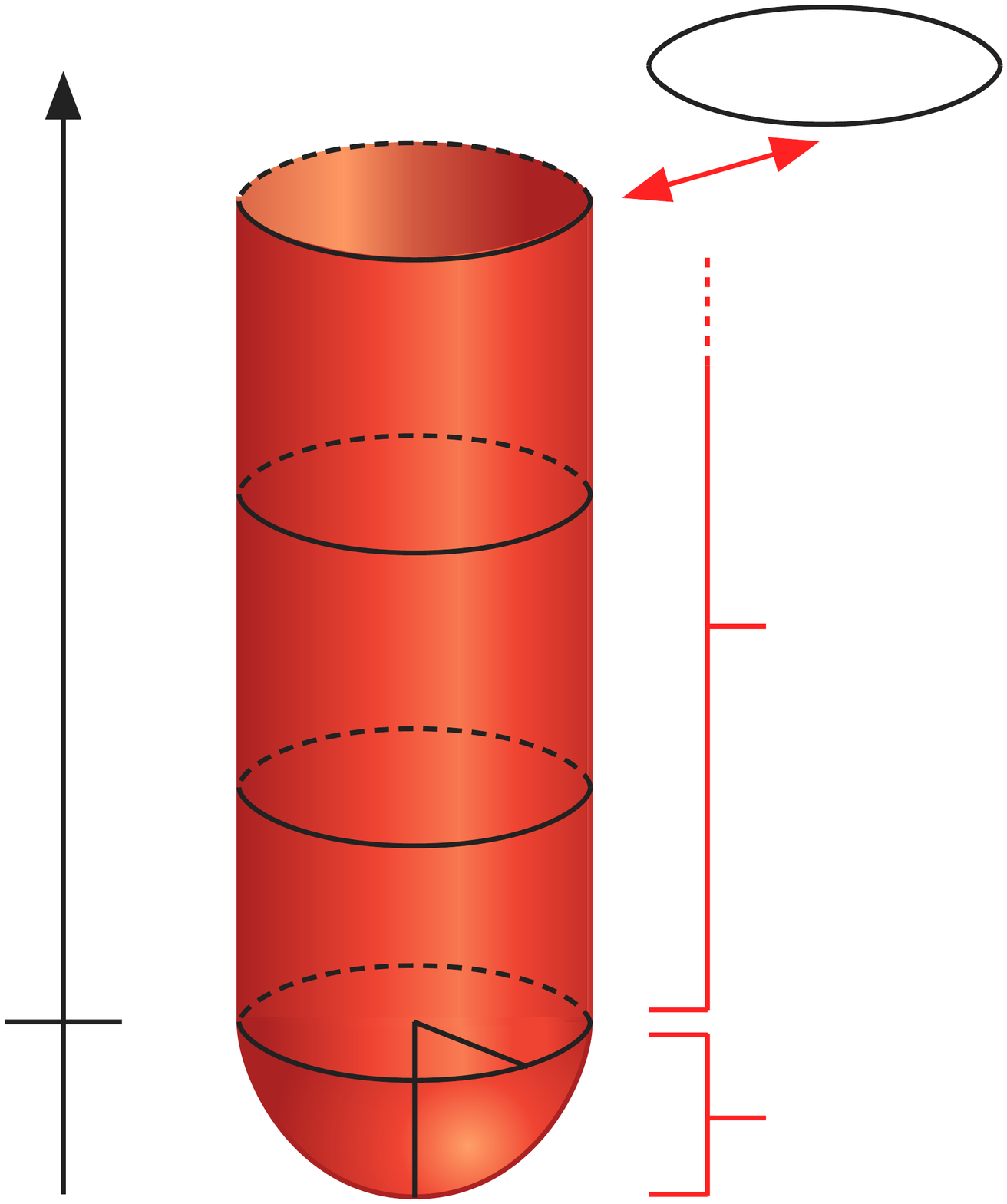}
 \put (5,88) {$y$} \put (0,24) {0} \put (56,18) {$S^3_{1/2}$} \put (56,50) {$S^2 \times \mathbbm{R}$} \put (52,85) {$=S^2$} \put (34,25) {$R$} \put (26,17) {$R$}
\end{overpic}
\caption{The geometry close to the tip of the throat.}\label{fig:The geometry}
\end{figure}

Let $y$ be the radial coordinate such that $y=0$ at the boundary of $S^3_{1/2}$ (see Figure \ref{fig:The geometry}).
Our starting point is the type IIB supergravity action in Einstein frame (see \cite{Polchinski:1998rr}, ch.~12.1):
\begin{equation}
S_{IIB}\supset \frac{1}{2\kappa_{10}^ 2}\left(\int d^{10}x \sqrt{-g^{10d}}R-\frac{g_s}{2}\int F_3\wedge *F_3-\frac{1}{2g_s}\int H_3 \wedge *H_3\right)\; ,
\end{equation}
where $F_3=dC_2$ and $H_3=dB_2$ are the three-form field-strengths and we have restricted ourselves to constant dilaton $e^{\phi}\equiv g_s$ and vanishing $C_0$.
We now expand:
\begin{equation}
C_2=\phi (x,y)\,\omega_2 \quad , \; x \in \mathbb{R}^{1,3}\; ,
\end{equation}
where we take $\omega_2$ to be the canonical volume-form of $S^2$ (normalized to $\omega_2=(Vol_{S^2})^{-1}\, *_2\, 1$). We now want to derive the effective 5d-action which we will then treat as an effectively 5d Randall-Sundrum-model \cite{RS-LargeMassHierarchy,RS-AlternativeComp} in Section \ref{5d-4d} (see \cite{ThroatasRandall-Sundrum} for the 5d-description of the throat).

First we derive the bulk-term. Thus we plug the above into the 10d-(Einstein-frame)-action $S_{IIB}\supset -\frac{g_s}{4\kappa_{10}^2} \int dC_2 \wedge * dC_2$ and get a bulk kinetic term
\begin{equation}
-\frac{g_s}{4\kappa_{10}^2}\int_{M_5} d^5 x \sqrt{-g^{5d}}\, \del_A \phi\, \del^A \phi \, \int_{S^3_{KS}}*_3\,1\,\underbrace{\int_{S^2} \omega \wedge *_{2} \,\omega}_{=(Vol_{S^2})^{-1}\equiv a_1} \quad,\; A\in \{0,...,4\}\; .
\end{equation}
Next let us calculate the contribution of the boundary $S^3_{1/2}$.
Since $S^2$ is trivial (e.g. at $y=0$, $S^2=\del S^3_{1/2}$) we have
\begin{equation}
\phi(x,y)= \int_{\del S^3_{1/2}} C_2=\int_{S^3_{1/2}} F_3\; ,
\end{equation}
with $F_3=dC_2$. Neglecting the warping, the lowest energy configuration is where the field-strength $F_3$ is equally distributed over $S^3_{1/2}$. Hence we make an ansatz
\begin{equation}
F_3=\gamma \, \omega_3 \quad , \; \gamma \in \mathbb{R}\; ,
\end{equation}
where $\omega_3$ is the canonical volume form of the three-sphere (i.e. $\omega_3=(Vol_{S^3})^{-1}\, *_3\, 1$). It follows that
\begin{equation}
\phi(x,0)=\int_{S^3_{1/2}} F_3= \frac{1}{2} \int_{S^3} F_3 = \frac{\gamma}{2} \quad
\rightarrow \quad F_3 = 2\phi(x,0)\; \omega_3\; .
\end{equation}
Plugging this into the 10d-action we get a boundary mass term
\begin{equation}
\label{mass-term-equ}
-\frac{g_s}{4\kappa_{10}^2}\int F_3 \wedge *\, F_3=-\frac{g_s}{4\kappa_{10}^2}\int_{\mathbb{R^{1,3}}} d^4 x \sqrt{-g^{5d}_{y=0}}\, 4\,\phi^2(x,0) \int_{S^3_{KS}}*_3\,1\, \underbrace{\int_{S^3_{1/2}}\omega_3 \wedge *_3 \, \omega_3}_{=\frac{1}{2}(Vol_{S_3})^{-1}\equiv a_2}\; ,
\end{equation}
where we have again neglected the effect of warping on $S^3_{1/2}$. Going over to a canonically normalized $5d$-field ($\frac{g_s}{2\kappa_{10}^2} a_1\,Vol_{S^3_{KS}}\,\phi^2\; \rightarrow\; \phi^2$) we get a 5d-action
\begin{equation}
\label{5d-action}
S_5=\int \sqrt{-g^{5d}}\,\{-\frac{1}{2} \del_A \phi \del^A \phi-\frac{1}{2}\Lambda \delta(y)\, \phi^2\} \quad \text{with}\quad \Lambda=4\frac{a_2}{a_1}=\frac{4}{\pi R}\; .
\end{equation}
Therefore the localized mass-term is essentially $\Lambda\sim R^{-1}$ where $R$ is the typical transverse size of the throat which in this case coincides with the length-scale over which the throat contracts.
\section{KK-Reduction on the Effective 5d-Throat and the 4d Action}
\label{5d-4d}
The 5d action derived in the previous section can now be reduced to an effective 4d action containing an infinite tower of 4d-KK-modes. We now treat the throat as an effectively 5-dimensional Randall-Sundrum-model \cite{RS-LargeMassHierarchy,RS-AlternativeComp,ThroatasRandall-Sundrum}. 

Consider the following 5d-metric \cite{RS-AlternativeComp,RS-LargeMassHierarchy}:
\begin{equation}
ds^2=e^{2ky}\eta_{\mu\nu}dx^{\mu}\otimes dx^{\nu}+dy^2\; .
\end{equation}
The 5d Lagrangian now reads
\begin{equation}
\mathcal{L}_5=e^{4ky}\{-\frac{1}{2}e^{-2ky}\del_{\mu}\phi\del^{\mu}\phi-\frac{1}{2}(\del_{y}\phi)^2-\frac{\Lambda}{2}\delta(y)\phi^2\}\; ,
\end{equation}
where 4d indices are contracted using $\eta=\text{diag}(-1,1,1,1)$. \par 
We now let $y$ take values on a strip of length $L$ choosing orbifold identification $y\cong -y$ and $y\cong y+2L$ (note that in this case we need to double $\Lambda$, that is $\Lambda=8/\pi R$, in order to get the physical boundary-condition for the 5d-field). \par 
Inserting a $4d$ plane wave ansatz $\phi(x,y)=e^{ipx} \chi(y)$ with $p^2=-m^2$ the equations of motion take the form
\begin{equation}
\label{eom}
-\del_y(e^{4ky}\del_y \chi)+\Lambda e^{4ky} \delta(y)\chi=e^{2ky} m^2 \chi\; ,
\end{equation}
which can be brought into the form of a 1d Schrödinger equation \cite{RS-AlternativeComp} (we follow explicitly \cite{QMPotential})
\begin{equation}
\label{SchroedingerEqu}
(-\del_z^2 + V(z))f(z)=E\,f(z)\; ,
\end{equation}
with $z\equiv e^{-ky}$, $\chi=z^{\frac{3}{2}}f$, $E=\frac{m^2}{k^2}$ and potential
\begin{equation}
\label{potential}
V(z)=\frac{15}{4}\frac{1}{z^2}+\frac{3+\Lambda/k}{z_{IR}}\delta(z-z_{IR})-\frac{3}{z_{UV}}\delta(z-z_{UV})\; ,
\end{equation}
where $z_{IR}=1$ and $z_{UV}=e^{-k L}$. 
\par Note that the delta-potentials come from enforcing the appropriate boundary conditions on $\chi$ (not on $f$). The general solution (a special case of the more general situation considered in \cite{GoldbergerWiseBulk}) now takes the form
\begin{equation}
\begin{split}
& f(z)=\sqrt{z}\left(A\,J_2\left(\frac{m}{k}z\right)+B\,Y_2\left(\frac{m}{k}z\right)\right)\quad,\; A,B\in \mathbb{C}\\
\Rightarrow\quad &\chi(y)=e^{-2ky}\left(A\,J_2\left(\frac{m}{k}e^{-ky}\right)+B\,Y_2\left(\frac{m}{k}e^{-ky}\right)\right) \; ,
\end{split}
\end{equation}
where $J_n$ and $Y_n$ are the Bessel functions of first and second kind respectively. \par 
From the form of the potential we immediately deduce the existence of a single (UV-) bound state and wave solutions of higher energy (mass) that are exponentially suppressed in the UV. Note that the bound state solution can be determined exactly in the case where $\Lambda=0$:
\begin{equation}
\begin{split}
f_0(z)=A\, z^{-\frac{3}{2}} + B\, z^{\frac{5}{2}} \;\Rightarrow \; \chi_0(z)=A+B\, z^4 \; ,
\end{split}
\end{equation}
which simplifies to $\chi=const.$ after imposing boundary conditions. This is of course the constant mode of zero mass which can be immediately read of from \eqref{eom}. \par 
The mass-condition follows from the two boundary conditions ($\del_y\chi(0)=\frac{\Lambda}{2}\chi(0)$ and $\del_y \chi(e^{-kL})=0$) and reads
\begin{equation}
J_1\left(\frac{m}{k}\right)+\frac{\Lambda}{2m}J_2\left(\frac{m}{k}\right)-\frac{J_1(\frac{m}{k}e^{-kL})}{Y_1(\frac{m}{k}e^{-kL})}\left(Y_1\left(\frac{m}{k}\right)+\frac{\Lambda}{2m}Y_2\left(\frac{m}{k}\right)\right)=0\; .
\end{equation}
We will now focus on the case $r_c\equiv\frac{1}{k}\ll L$ (which is the interesting case of strong warping). For the bound-state solution we expect a small mass ($m\ll k$) for which we can use the small argument approximations of the Bessel-functions 
\begin{equation}
\begin{split}
& J_1(x)=\frac{x}{2}+\mathcal{O}(x^3)\quad J_2(x)=\frac{x^2}{8}+\mathcal{O}(x^4)\\
& Y_1(x)=-\frac{2}{\pi x}+\mathcal{O}(x) \quad Y_2(x)=-\frac{4}{\pi x^2}+\mathcal{O}(x^0)\; ,
\end{split}
\end{equation}
to arrive at
\begin{equation}
\label{smallmass}
\frac{m_0}{k}=\left(\frac{k}{\Lambda}+\frac{1}{8}\right)^{-\frac{1}{2}}e^{-kL}\; .
\end{equation}
Remarkably this mass is exponentially suppressed by the warp factor (thereby a posteriori justifying our small argument approximation). It is crucial to realize that this is not the usual hierarchy induced by warping in Randall-Sundrum models \cite{RS-LargeMassHierarchy} but is rather a suppression 'on top of that' since our metric conventions are such that $g^{IR}_{\mu\nu}\equiv g_{\mu\nu}(y=0)=\eta_{\mu\nu}$. 
\par
\begin{figure}
\begin{tabular}{cc}
\includegraphics[scale=0.55,keepaspectratio]{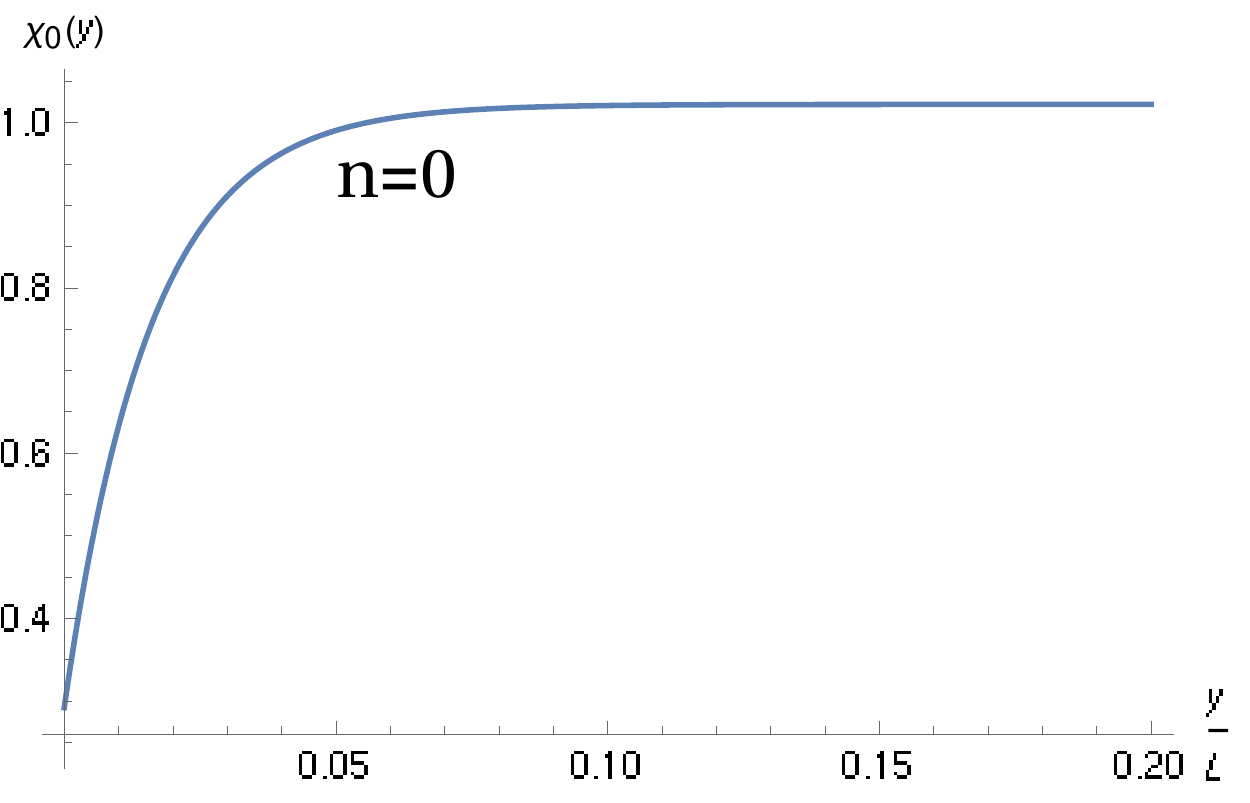} & \includegraphics[scale=0.55,keepaspectratio]{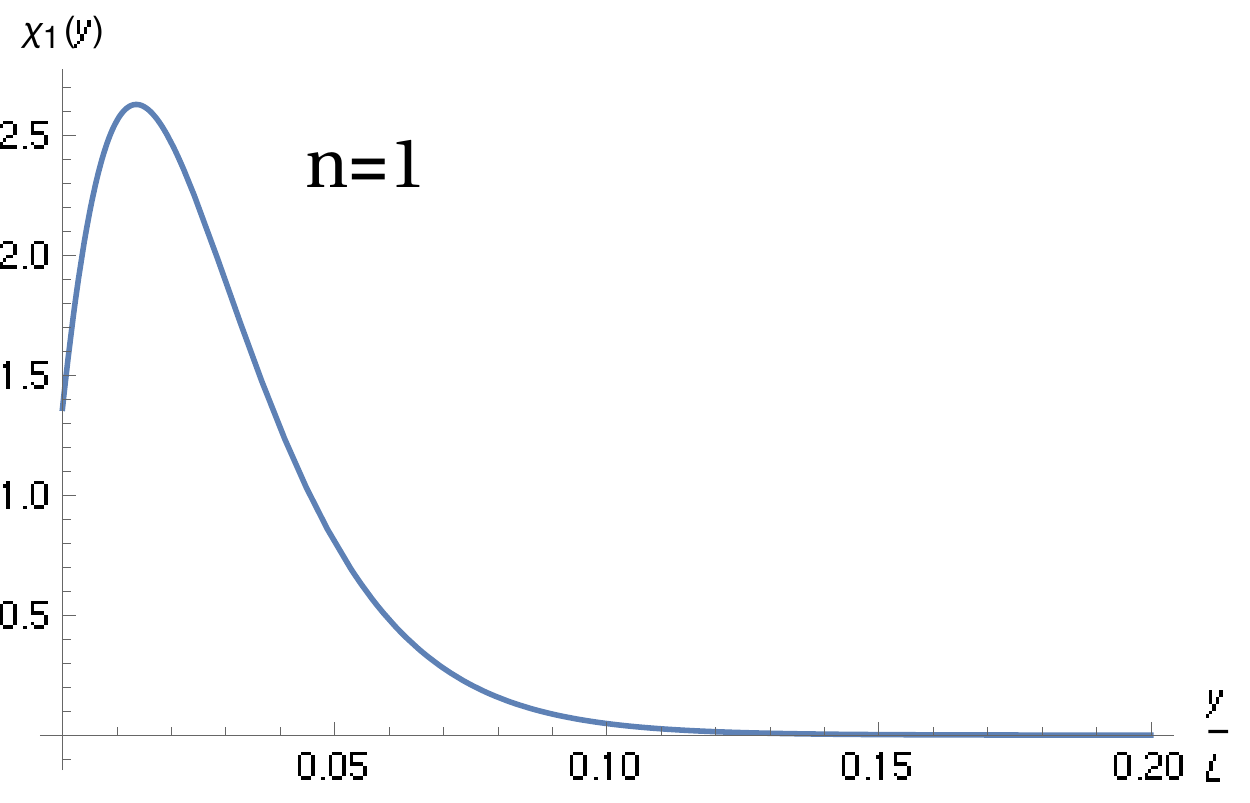} \\ 
\includegraphics[scale=0.55,keepaspectratio]{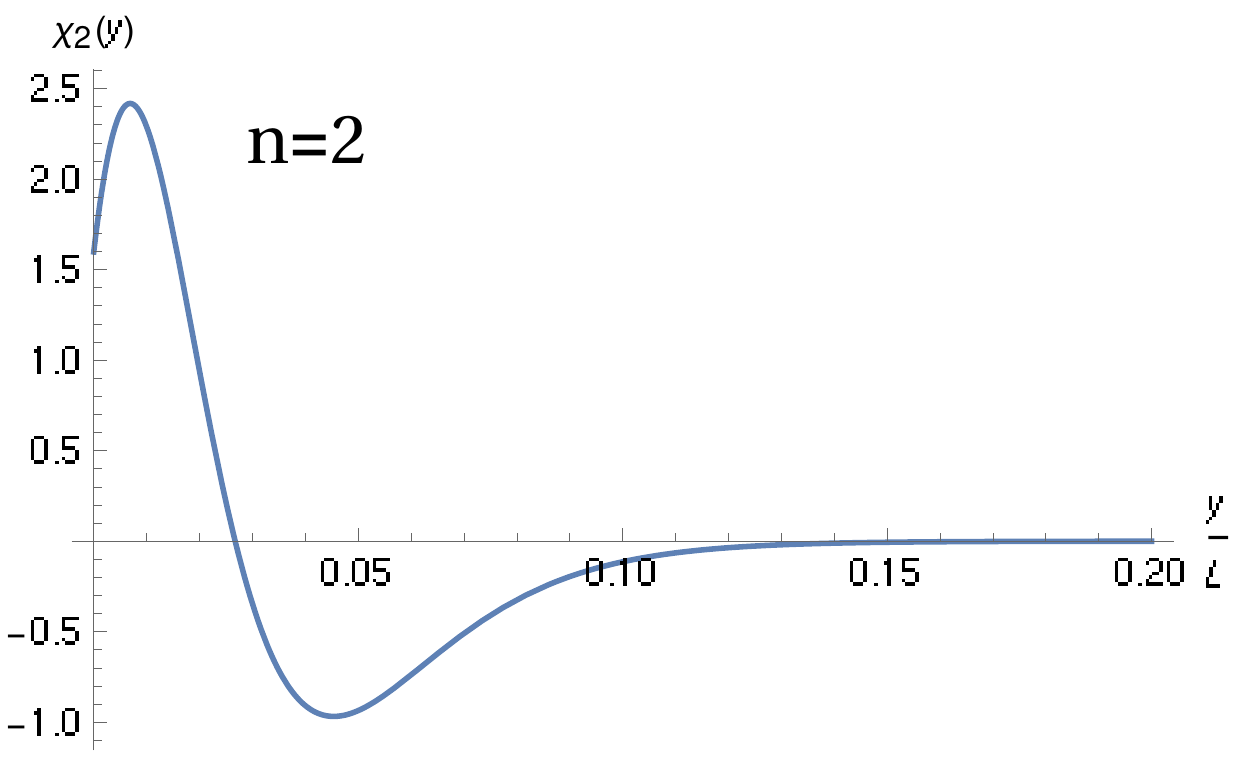} & \includegraphics[scale=0.55,keepaspectratio]{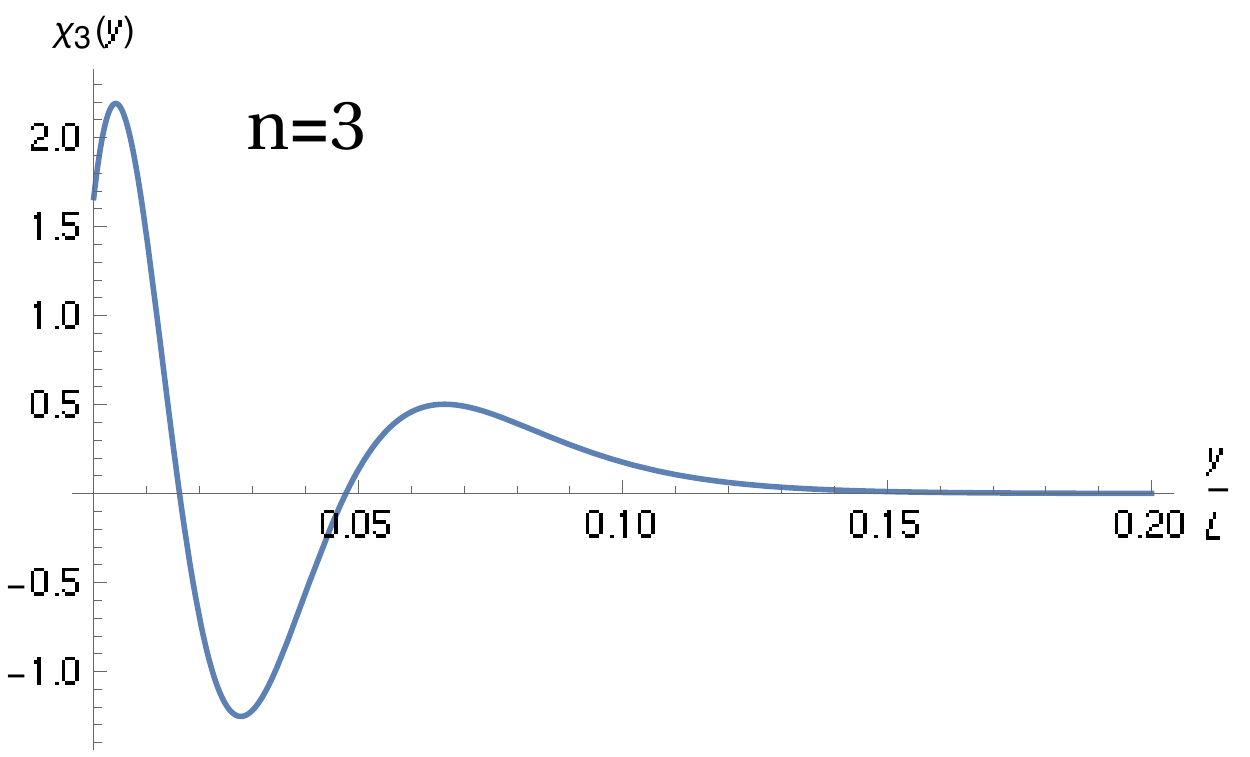} \\ 
\end{tabular} 
\caption{The bound state and first excited solutions with $kL=5 \pi$ and $\Lambda=100 \pi/L$ ($r_c/L\approx 0.06$), plotted in the range $0\leq y \leq 0.2 \,L$}
\label{Plots}
\end{figure}

 The zero-mode profile takes the following form:
\begin{equation}
\label{boundmode}
\chi_0(y)\propto \left(1-\frac{1}{8}\frac{1}{k/\Lambda+1/8}e^{-4ky}\right)\; .
\end{equation}
The higher KK-modes (with $1\lesssim\frac{m}{k}\ll e^{kL}$) are obtained by noting that $J_1/Y_1(x)=-\frac{\pi}{4}x^2+\mathcal{O}(x^4)$ such that the mass condition is approximately
\begin{equation}
J_1\left(\frac{m}{k}\right)+\frac{\Lambda}{2m}J_2\left(\frac{m}{k}\right)=0\; .
\end{equation}
The solutions interpolate between the zeros of the two Bessel-functions ($j_{1,n}$ and $j_{2,n}$), that is
\begin{equation}
\begin{split}
& m\ll \Lambda:\quad \frac{m_n}{k}\approx j_{2,n}\\
& m\gg \Lambda:\quad \frac{m_n}{k}\approx j_{1,n}
\end{split}
\end{equation}
and asymptotically (that is $m_n\gg k,\Lambda$) 
\begin{equation}
\label{KK-mass}
m_n=\pi \frac{n}{r_c} \; ,
\end{equation}
which are the usual KK-masses but with $L$ replaced by the curvature radius $r_c\equiv k^{-1}$. The bound-state and the first excited states are plotted in Figure \ref{Plots}.
\par
Using the 4d-Planck mass \cite{RS-LargeMassHierarchy}
\begin{equation}
M^2_{pl}=\frac{M^3_5}{k}\left(e^{2kL}-1\right)\overset{kL\gg 1}{\approx} \frac{M^3_5}{k}e^{2kL}\; ,
\end{equation}
one immediately sees the double exponential suppression of the bound-mode:
\begin{equation}
\begin{split}
&\frac{m^2_0}{M^2_{pl}}\approx \frac{k^3}{M^3_5}(k/\Lambda+1/8)^{-1}e^{-4kL}\propto e^{-4kL}\\
&\frac{m^2_n}{M^2_{pl}}\approx \frac{k^3}{M^3_5}\pi^2 n^2 e^{-2kL}\propto e^{-2kL} \quad \forall\; n\neq 0\; ,
\end{split}
\end{equation}
Note that this agrees with the expression for the axion-potential in equation (4.76) of \cite{GravitywavesLinearInflation} where the potential comes from the NS5-DBI-action.\par
This behavior could have already been anticipated from the form of the potential \eqref{potential}: The bound-state-solution approaches a constant in the UV while the positive delta-potential in the IR leads to a dip in the IR. It therefore gets its mass from the IR while its kinetic term lives in the whole bulk (concerning the kinetic term arguments along these lines have already been given in \cite{GravitywavesLinearInflation}, Sec. 4.3.2). This leads to the already mentioned `double'-suppression. The higher KK-modes are the solutions to Schrödinger's equation \eqref{SchroedingerEqu} that oscillate in the IR-region $0<y\lesssim r_c$ and fall off exponentially towards the UV due to the $\sim 1/z^2$-term in the potential \eqref{potential}. This leads to the modified KK-mass-formula \eqref{KK-mass}. \par 
Note furthermore that $m_0$ (more precisely its upper bound) is not particularly sensitive to the value of $\Lambda$:
\begin{equation}
0\leq m_0 \leq \sqrt{8}k e^{-kL}\quad \Rightarrow\quad
0\leq\frac{m_0}{m_1}\leq (j_{1,1})^{-1}\sqrt{8}\, e^{-kL}\; ,
\end{equation}
where $j_{1,1}\approx 3.8317...$ is the first zero of $J_{1}$. The 4d-effective action is
\begin{equation}
S=\int_{\mathbb{R}^{1,3}}d^4 x\, \sum_{n=0}^{\infty}\{-\frac{1}{2}\del_{\mu}\phi_n \del^{\mu}\phi_n-\frac{1}{2}m_n^2 \phi^2_n\}\; .
\end{equation}
Let us pause here and highlight what we have found:\par 

The lightest KK-mode of the RR-2-form $C_2$ on the KS-throat with trivial $S^2$-cycle is exponentially lighter than the next higher mode in the case of strong warping.
This makes it an ideal candidate for single-field chaotic inflation since we can safely ignore the higher modes.
\section{Simple Consistency Checks from Energetics and Mass Scales}
\label{Checks}
\subsection{Energy Density at the boundary of $S^3_{1/2}$}
It is important to check that the energy density at $y=0$ on the cylinder is the same (at least up to $\mathcal{O}(1)$-factors) as the one on $S^3_{1/2}$. \par 
On the cylinder we have
\begin{equation}
S_{cyl}=-\int d^5 x \int_{S^3_{KS}}*_{3}1\int_{S^2}*_{2}1\sqrt{-g^{5d}}\underbrace{\frac{g_s}{4\kappa_{10}^2}(Vol_{S^2})^{-2} (\del_y \phi)^2}_{\equiv \eps_{cyl}}\; ,
\end{equation}
while on $S^3_{1/2}$ we have
\begin{equation}
S_{S^3_{1/2}}=-\int_{\mathbb{R}^{1,3}} d^4 x \int_{S^3_{KS}}*_{3}1\int_{S^3_{1/2}}\sqrt{g^{S^3_{1/2}}}\sqrt{-g^{4d}}\underbrace{\frac{g_s}{4\kappa_{10}^2}(Vol_{S^3})^{-2} 4(\phi)^2}_{\equiv\eps_{S^3_{1/2}}}\; ,
\end{equation}
which implies that
\begin{equation}
\frac{\eps_{cyl}}{\eps_{S^3_{1/2}}}= \frac{1}{4}\underbrace{\left(\frac{\del_y \phi(0)}{\phi(0)}\right)^2}_{\equiv (\Lambda/2)^2}\underbrace{\left(\frac{Vol_{S^3}}{Vol_{S^2}}\right)^2}_{\equiv (\frac{\pi}{2}R)^2}=1\; .
\end{equation}
Therefore the energy-densities are exactly the same. Note that this were also true if we had chosen any other eigen-mode of the 5d-Laplacian since the identity $\Lambda=\del_y \chi_n(0)/\chi(0)$ is simply the boundary condition for the $y$-profile of \textit{any} mode $\chi_n$.
\subsection{Inflaton Energy Density vs $F_3$-Flux Energy Density}
Since we have a model of single-field large field inflation we have to make sure that the field excursion of the inflaton does not back-react in a way that destabilizes the throat. \par 
The flux energy density can be calculated from the type IIB Supergravity action
\begin{equation}
S_{RR}\supset -\frac{g_s}{4\kappa_{10}^2}\int F_3\wedge * F_3 \quad\text{with}\quad F_3 = (2\pi)^2\alpha'M \omega_3+... \; ,
\end{equation}
where $\omega_3$ is the appropriately normalized volume form on $S^3_{KS}$ and $M$ is the $F_3$-flux on $S^3_{KS}$ stabilizing the throat. Ellipsis indicate terms that integrate to zero over $S^3_{KS}$. \par Using $\kappa_{10}^2=\frac{l_s^8}{4\pi}$ (where $l_s=2\pi \sqrt{\alpha'}$) and $\int \omega_3\wedge * \omega_3=(Vol_{S^3})^{-2}\int d^{10}x \sqrt{-G^{10d}}$ this yields the local $10d$ energy density
\begin{equation}
\eps^{10d}_{KS}=\frac{g_sM^2}{4\pi^3 R^6 l_s^4}=\frac{2^4 \pi^3}{l_s^{10}}(Mg_s^{1/2} )^{-1}\; ,
\end{equation}
where $R$ is the radius of the $S^3_{KS}$ (which we identify with the $S^3_{1/2}$ radius). In the second step we have used that $R^2=Mg_s^{1/2} (l_s/2\pi)^2$ in the KS-region\footnote{This differs from the length-scale given in \cite{KlebanovStrassler} by a factor of $\sqrt{g_s}$ as we are working in Einstein-frame.}.\par 
The inflaton energy density (using equations \eqref{mass-term-equ},\eqref{5d-action} and the explicit form of the bound mode \eqref{boundmode}) is given by
\begin{equation}
\begin{split}
&\eps_{\phi}^{10d}=\frac{g_s\phi^2}{\kappa_{10}^2(Vol_{S^3})^2}=\frac{2\phi_{5d-can.}^2(Vol_{S^2})}{(Vol_{S^3})^3}=\frac{2 \alpha^2}{\pi^5}\frac{M_5^3}{R^7}\frac{k/\Lambda}{k/\Lambda+1/8}\\
& \leq \frac{2\alpha^2}{\pi^5}\frac{M_5^3}{R^7}= \frac{2^6\alpha^2}{\pi}\frac{1}{R^2 l_s^8}=\frac{2^{8}\pi}{l_s^{10}}\alpha^2(Mg_s^{1/2})^{-1}\; ,
\end{split}
\end{equation}
where $\alpha$ measures the 4d field excursion in 4d-Planck units (equivalently the 5d excursion in 5d-Planck units). The ratio of the densities therefore satisfies
\begin{equation}
\eps_{\phi}/\eps_{KS}\leq\frac{16}{\pi^2}\alpha^2\sim \alpha^2\; .
\end{equation}
Therefore in the interesting regime of large field, $\alpha^2\gg 1$, the back-reaction on the ambient geometry cannot be neglected. The full non-linear equations of type IIB Supergravity have to be considered to quantify this back-reaction.
\section{The Ultra-light Mode in the KS background}
\label{section:UltraLightModeKS}
In the following we would like to describe the ultra-light mode in the full Klebanov-Strassler (KS) geometry in order to address questions of back-reaction. Since back-reaction effects take place at the tip of the throat only, where the metric is known, this can be done explicitly.\par 
To this end we will specify an explicit ansatz that describes our mode in the KS-background and derive the equations of motion. Obtaining the full solution is an involved numerical task, that will be left for future research.
\subsection{A Simple Prescription for Obtaining the Back-reacted Potential}
\label{Non-linear backreaction}
Before turning to the relevant equations of motion, let us discuss how the effective back-reacted potential in $4d$ can be obtained without having to solve complicated time-dependent equations of motion. As we will see, the effective $4d$ potential can be efficiently extracted by considering static and homogeneous field profiles $\phi=\phi(\tau)$.\par
Let us parameterize the effective $5d$ action as follows,
\begin{equation}
\label{S_obtainpotential}
\begin{split}
S_{\text{eff},5d}^0[\phi]=&\int_{4d}d^4 x \int d\tau\,\mathcal{L}_{\text{eff}}^0(\phi,\del_{\tau}\phi,\tau)\, ,\\ \mathcal{L}_{\text{eff}}^0(\phi,\del_{\tau}\phi,\tau)=&-\frac{M_5^3}{2}X(\tau)\sqrt{g_{\tau\tau}}\left(e^{2\mathcal{A}}\eta^{\mu\nu}\del_{\mu} \phi\del_{\nu}\phi+e^{4\mathcal{A}}g^{\tau\tau}(\del_{\tau}\phi)^2\right)+\mathcal{L}_{\text{int}}(\phi,\del_{\tau}\phi,\tau)\, ,
\end{split}
\end{equation}
where we use the for now arbitrary radial coordinate $\tau$ which does not necessarily measure physical distances (i.e. $g_{\tau\tau}$ need not be unity).
Here the function $X(\tau)$ parameterizes the varying volume of $T^{1,1}$ and its $2$-cycle which appear in the dimensional reduction from $10d$ to $5d$ and $e^{2\mathcal{A}}$ is the warp factor. We have not written out explicitly any terms beyond quadratic order in $\phi$. These are included in $\mathcal{L}_{\text{int}}$ which we assume to take significant values only near the IR. Clearly there is no static homogeneous solution to the equations of motion with the boundary conditions $\phi(0)=0=\del_{\tau}\phi(\tau_{UV})$ other than the trivial solution $\phi=0$. This is expected as we know that the lowest lying mode obtains a non-vanishing potential from the $4d$-perspective and can hence not be static. However, if a source $j$ is inserted at the UV boundary,
\begin{equation}
\mathcal{L}_{\text{eff}}^0\longrightarrow\mathcal{L}_{\text{eff}}^1=\mathcal{L}_{\text{eff}}^0+j\cdot \phi(\tau)\delta(\tau-\tau_{UV})\, ,
\end{equation}
a non-trivial solution is obtained as the UV-boundary conditions are altered to \begin{equation}
\label{source1}
M_5^3Xe^{4\mathcal{A}}\sqrt{g^{\tau\tau}}\del_{\tau}\phi{\Big|}_{\tau=\tau_{UV}}=j\, .
\end{equation}
For given source $j$ there is hence a non-trivial static profile $\phi(\tau)$ that solves the (non-linear) equations of motion. Intuitively the source $j$ sets the field excursion by applying a restoring force against the potential slope. Let us parameterize the field-excursion by the value $\phi_{UV}\equiv \phi(\tau_{UV})$. Then to each value of the source $j$ there is an associated field excursion $\phi_{UV}$ and (on-shell) we can hence interpret the source $j$ as a function of the field excursion,
\begin{equation}
\label{source_onshell}
j=j(\phi_{UV})\, .
\end{equation}
It should be noted that in order to obtain this function $j(\phi_{UV})$ explicitly, the non-linear equations of motion have to be solved numerically. The function $j(\phi_{UV})$ is then a complicated non-linear function that is known only numerically.\par 
Let us now change perspective and analyze the same problem from the effective $4d$ point of view. The $4d$ action is
\begin{equation}
S=\int d^4x \left(-\frac{f^2}{2}(\del \phi_{UV})^2-V(\phi_{UV})+j\cdot \phi_{UV}\right)\, ,
\end{equation}
with axion decay constant $f$ and potential $V(\phi_{UV})$. At this stage, the potential $V(\phi_{UV})$ is unknown. Again, there is a static configuration at field excursion $\phi_{UV}$ if
\begin{equation}
\label{source2}
V'(\phi_{UV})=j\, .
\end{equation}
Because both the $5d$ point of view as well as the effective $4d$ point of view should give the same answer, the potential $V(\phi_{UV})$ can be inferred by comparing \eqref{source1} with \eqref{source2}.
Finally, we have obtained the desired simple prescription to read off the effective $4d$-potential from a static numerical solution of the non-linear bulk equations of motion with boundary conditions $\phi(0)=0$ and $\phi(\tau_{UV})=\phi_{UV}$\footnote{Because the $4d$-potential comes from field-gradients that are localized in the IR, it is indifferent to the details of the UV-geometry. The fact that we determine the potential by an expression that is evaluated in the UV does not imply that the final result is sensitive to the details of the UV-geometry.}. The crucial advantage is that there is no need for an explicit dimensional reduction of the higher-dimensional action to $4d$.\par 
Let us now specify to the case of strong warping, approximately constant field-profile $\phi(\tau)$ and $g_{\tau\tau}\approx const \sim k^{-2}$ at large $\tau$, where $k$ is the inverse curvature radius of the effective $5d$ geometry. Then the kinetic term is dominated by the UV-region and it follows that 
\begin{equation}
f^2\approx M_5^3 X(\tau)\sqrt{g_{\tau\tau}}e^{2\mathcal{A}}|_{\tau=\tau_{UV}}\, ,
\end{equation}
where we have dropped overall factors of $\mathcal{O}(1)$ that are not affected by back-reaction. In the case of the KS-throat one has that  $k^{-2}\sim g_s M \alpha'$ and we call $m_{wKK}^2\equiv (g_s M \alpha')^{-1}$ the warped KK-scale which is the mass-scale of KK-modes that are localized at the tip of the throat. Then it follows that the effective potential in $4d$ can be expressed as
\begin{equation}
\label{4dbackreactedpotential}
\begin{split}
\frac{V(\phi_c)}{M_{pl}^4}&=\frac{m_{wKK}^2}{M_{pl}^2}\cdot\underbrace{\int_{0}^{\phi_c}\frac{d\tilde{\phi_c}}{M_{pl}}\,\,\cdot \frac{\tilde{\phi_c}}{M_{pl}}\cdot \gamma^2\left(\frac{\tilde{\phi_c}}{f}\right)}_{\equiv I(\phi_c)}\,\,\, ,\\
\text{where}\quad \gamma^2(\phi_{UV})&\equiv\left(e^{2\mathcal{A}}\frac{\del_{\tau}\phi(\tau)}{\phi(\tau)}\right)_{\tau=\tau_{UV}\, ;\, \phi(\tau_{UV})=\phi_{UV}}\, ,
\end{split}
\end{equation}
with canonically normalized $4d$ field $\phi_c$ and axion-decay constant $f$.\par 
The factor of $\frac{m_{wKK}^2}{M_{pl}^2}$ is the suppression enjoyed by an IR-brane field (in the spirit of RS1 \cite{RS-LargeMassHierarchy}), while the integral $I(\phi_c)$ quantifies the extra suppression that only the ultra-light mode enjoys and corrections due to back-reaction at large field excursion. 
For small field excursions one recovers the quadratic potential of Section \ref{5d-4d}, $V=\frac{1}{2}m_0^2 \phi_c^2$, with squared mass \begin{equation}
m_0^2\equiv m_{wKK}^2\lim\limits_{\phi_{UV}\longrightarrow 0}\gamma^2(\phi_{UV})\, ,
\end{equation} 
which corresponds to a source $j$ that is linear in the field excursion $\phi_{UV}$.
\subsection{The Type IIB Equations of Motion}
Having learned how to extract the effective potential from a solution to the equations of motion we now derive the explicit equations of motion that need to be solved eventually. We start with the \textit{String frame} equations of motion and Bianchi-identities of type IIB Supergravity (for now omitting the Einstein equations):
\begin{equation}
\label{SUGRAeomC2}
d*\tilde{F}_3=F_5\wedge H_3\,, \quad d\tilde{F}_3=-F_1\wedge H_3 ,
\end{equation}
\begin{equation}
\label{SUGRAeomB2}
d(e^{-2\phi}*H_3)=g_s^2(F_1\wedge *\tilde{F}_3-F_5\wedge \tilde{F}_3)\,,\quad dH_3=0\, ,
\end{equation}
\begin{equation}
\label{SUGRAeomC4}
dF_5=H_3\wedge \tilde{F}_3\,,\quad
F_5=*F_5\, ,
\end{equation}
\begin{equation}
\label{SUGRAeomC0}
d*F_1=-H_3\wedge*\tilde{F}_3\, , \quad dF_1=0\, ,
\end{equation}
\begin{equation}
\label{SUGRAeomPhi}
d(e^{-2\phi}*d\phi)=\frac{1}{16}(e^{-2\phi}H_3\wedge *H_3-g_s^2\tilde{F}_3\wedge *\tilde{F}_3 )-\frac{1}{8}g_s^2F_1\wedge *F_1\, .
\end{equation}
In practice we will work with $F_1=d C_0$, $H_3=dB_2$ and $\tilde{F}_3=F_3-C_0\wedge H_3$ and specify an ansatz for $F_3$, $B_2$ and $C_0$ such that $dF_3=0$. We will not work with a four-form potential and specify an ansatz directly for $F_5$. Furthermore we redefine the dilaton $\Phi$ as $e^{\Phi}\equiv g_s e^{\phi}$ and define $g_s$ to be the value that $e^{\Phi}$ approaches in the UV.
\subsection{The Ansatz and Boundary Conditions} 
Far away from the tip the mode we are looking for should approach $\delta C_2=f(\tau)\omega_2$ with harmonic two-form $\omega_2$ and nearly constant radial profile $f(\tau)$. Thus, the field strength of our mode and the corresponding back-reaction are negligible in this region. Near the tip, however, the KS solution has to be generalized in order to allow for back-reaction. Because the harmonic two-form does not break any of the global symmetries of the KS solution we can extend the KS ansatz in the most generic way that is compatible with these symmetries. \par 
The deformed conifold can be parameterized by a radial coordinate $\tau$ and a set of Euler angles $\{\phi_1,\theta_1,\phi_2,\theta_2,\psi\}$ that parametrize the $5d$ space $T^{1,1}$. Its Kähler and Ricci-flat metric is invariant under $G=SU(2)\times SU(2)$, as well as a discrete $Z_2$ that interchanges the Euler angles $(\phi_1,\theta_1)\leftrightarrow(\phi_2,\theta_2)$ \cite{KlebanovStrassler, KlebanovRemarks}. The metric of $T^{1,1}$ is usually expressed in terms of a set of one-forms\footnote{These are for example given in eq. $(4)$ of \cite{KlebanovStrassler}.} $\{g^i\}_{i=1}^5$ and it can be shown that the continuous symmetry group of the deformed conifold acts like simultaneous $2d$ orthogonal transformations on the vectors $v_1\equiv(g^1,g^2)^T$ and $v_2\equiv(g^3,g^4)^T$ while it has trivial action on $g^5$ \cite{Jakob-Masterthesis}. As a result, the following $2$-forms and symmetric $2$-tensors are invariant under the symmetries of the deformed conifold\footnote{It can be shown that they actually span the space of invariant $2$-tensor fields. $v_1^i\wedge v_2^j\eps_{ij}$ and $v_1^iv_2^j\delta_{ij}$ are also invariant under the continuous symmetries of the deformed conifold, but are odd under the $Z_2$.}:
\begin{equation}
\begin{split}
g^1\wedge g^2=\frac{1}{2}v_1^i \wedge v_1^j \eps_{ij}\, ,\quad
g^3\wedge g^4=\frac{1}{2}v_2^i \wedge v_2^j \eps_{ij}\, ,\quad
g^1\wedge g^3+g^2\wedge g^4=v_1^i\wedge v_2^j \delta_{ij}\, ,
\end{split}
\end{equation}
\begin{equation}
\label{T11metric}
(g^5)^2\, ,\quad (g^1)^2+(g^2)^2=v_1^iv_1^j\delta_{ij} \, , \quad(g^3)^2+(g^4)^2=v_2^iv_2^j\delta_{ij}\, ,\quad g^1g^4-g^2g^3=v_1^iv_2^j \eps_{ij}\, .
\end{equation}
Here, $g^i g^j\equiv \frac{1}{2}(g^i\otimes g^j+g^j\otimes g^i)$.\par 
Including the radial direction parameterized by the coordinate $\tau$, one may further allow for a term proportional to $d\tau \,g^5$ in the $6d$ metric since $g^5$ is invariant under the symmetries of the deformed conifold. This leads to the $6d$ metric
\begin{equation}
\label{6dmetric}
\begin{split}
ds^2_{6}=&a^2(\tau) (g^5)^2+b^2(\tau)((g^1)^2+(g^2)^2)+c^2(\tau)((g^3)^2+(g^4)^2)\\
&+d(\tau)(g^1g^4-g^2g^3)
+m(\tau)g^5 d\tau+n^2(\tau)d\tau^2\, ,
\end{split}
\end{equation}
with radial functions $a,b,c,d,m,n$. Demanding that the above is Kähler and Ricci-flat leads to the deformed conifold metric which in particular features $d=m=0$. Because we are interested in breaking the ISD-property of the KS solution we have no reason to assume $d=m=0$. However, \eqref{6dmetric} has a gauge redundancy that remains to be fixed: Under the re-parametrization $\psi \longrightarrow \psi +\lambda(\tau)$, one has that $g^5\longrightarrow g^5+\lambda'(\tau)d\tau$. Under such a re-parametrization the $6d$ metric is not invariant but again takes the form of \eqref{6dmetric} with different coefficients. In particular it can be checked that
\begin{equation}
\begin{pmatrix}
b^2-c^2\\
d
\end{pmatrix}\longrightarrow
\begin{pmatrix}
\cos(\lambda) & -\sin(\lambda)\\
\sin(\lambda) & \cos(\lambda)
\end{pmatrix}\cdot
\begin{pmatrix}
b^2-c^2\\
d
\end{pmatrix}\, .
\end{equation}
A non-vanishing function $d(\tau)$ can thus be gauged away by a suitable re-parametrization and we fix the gauge by setting $d(\tau)\equiv 0$.\par
Consequently we choose the ansatz 
\begin{equation}
\begin{split}
ds^2=\eta_{ab}\Theta^a\otimes \Theta^b\quad a,b\in\{x^{0},...,x^{3},\tau,5,1,...,4\}\, ,\quad
\eta_{ab}=\text{diag}(-1,1,...,1)_{ab}\, ,
\end{split}
\end{equation}
for the $10d$ metric. The zehnbein one-forms are
\begin{equation}
\label{zehnbein}
\begin{split}
&\Theta^{x^{\mu}} =A(\tau)dx^{\mu}, \quad \mu\in\{0,...,3\}\, ,\\
&\Theta^{\tau}=(g_s M \alpha')^{1/2}D(\tau)d\tau\, ,\quad \Theta^{5}=(g_s M \alpha')^{1/2}D(\tau)\left(g^5+E(\tau)d\tau\right)\, ,\\
&\Theta^{1}=(g_s M \alpha')^{1/2}B(\tau)g^1\, ,\quad
\;\Theta^{3}=(g_s M \alpha')^{1/2}C(\tau)g^3\, ,\\
&\Theta^{2}=(g_s M \alpha')^{1/2}B(\tau)g^2\, ,\quad \;\Theta^{4}=(g_s M \alpha')^{1/2}C(\tau)g^4\, ,
\end{split}
\end{equation}
with radially varying functions $A,B,C,D,E$\footnote{Allowing for $\Theta^{\tau}\propto L(\tau)d\tau$ with a generic function $L\neq D$ is seemingly more general but can be brought to the form $L=D$ by a suitable re-parameterization of the radial coordinate $\tilde{\tau}=\tilde{\tau}(\tau)$.}. One can check that this choice of zehnbein one-forms reproduces all the terms in \eqref{6dmetric} in a sufficiently general way.\par
We generalize the KS-ansatz to $F_3=\frac{M\alpha'}{2}g^5\wedge g^3\wedge g^4+d(\delta C_2)$ and $H_3=dB_2$ with
\begin{equation}
\begin{split}
\delta C_2=\frac{M\alpha'}{2}\left(f(\tau)g^1\wedge g^2+g(\tau)g^3\wedge g^4+F(\tau)(g^1\wedge g^3+g^2\wedge g^4)\right)\, ,\\
B_2=\frac{g_s M \alpha'}{2}\left(j(\tau)g^1\wedge g^2+k(\tau) g^3\wedge g^4-b(\tau) (g^1\wedge g^3+g^2\wedge g^4)\right)\, .
\end{split}
\end{equation}
It then follows by virtue of equation \eqref{SUGRAeomC4} that
\begin{equation}
\label{lC2}
\begin{split}
F_5=(1+*)\mathcal{F}_5\, ,\quad \mathcal{F}_5&=g_s\left(\frac{M \alpha'}{2}\right)^2l(\tau)g^1\wedge g^2 \wedge g^3 \wedge g^4 \wedge \left(g^5+Ed\tau\right)\, ,
\\
\text{with}\quad l(\tau)&=j(1-F)+k F+b(g-f)\, .
\end{split}
\end{equation}
Furthermore we allow for radial profiles of the axio-dilaton in a convenient parametrization
\begin{equation}
C_0=C_0(\tau)\equiv c(\tau)/g_s\, ,\quad \phi=\phi(\tau)\, .
\end{equation}
The IR boundary conditions are
\begin{equation}
\label{pformboundaryconditions}
\begin{split}
j(0)=k(0)=b(0)=g(0)=f(0)=F(0)=
\phi'(0)=c'(0)=0\, ,
\end{split}
\end{equation}
for the axio-dilaton and $p$-form fields, while we choose to parametrize the metric function $B$ by $B(\tau)\equiv\tanh(\tau)\mathcal{B}(\tau)$
and set
\begin{equation}
\label{metricboundaryconditions}
A'(0)=\mathcal{B}'(0)=C'(0)=D'(0)=0\, .
\end{equation}
Because the harmonic two-form of $T^{1,1}$ is proportional to $g^1\wedge g^2+g^3\wedge g^4$ we define the field excursion $\psi$ by imposing the UV boundary condition
\begin{equation}
\label{fieldexcursionboundarycondition}
f(\tau_{UV})=\psi=g(\tau_{UV})\, .
\end{equation}
Furthermore all other functions are required to take their KS-values. The set of IR boundary conditions \eqref{pformboundaryconditions} follows from demanding that field strengths be finite at the tip and \eqref{metricboundaryconditions} is a consequence of demanding the tip-topology to still be an $S^3$. 
The axio-dilaton is stabilized in the UV by ISD-fluxes on other cycles that are not relevant to the local KS-throat and we implement this by demanding that
\begin{equation}
\label{axio-dilaton_fieldexcursion}
c(\tau_{UV})=0=\phi(\tau_{UV})\, ,
\end{equation}
which corresponds to setting the UV-value of the axio-dilaton to $ig_s^{-1}$.\par
It should be noted that the IR boundary conditions that we have stated are conceptually very different from the UV-boundary conditions: The IR-boundary conditions are an unambiguous consequence of demanding that the solution gives rise to a smooth geometry with finite field-strengths at the tip. In contrast, the UV-boundary conditions are set such that on the $\tau=\tau_{UV}=const$ sub-manifold our solution coincides with the KS-solution, except with a non-vanishing Wilson 'line' of $C_2$ turned on (corresponding to $f(\tau_{UV})=g(\tau_{UV})\neq 0$). Because an exactly constant profile $f=g=const$ has vanishing field-strength and hence does not contribute in the equations of motion, we expect this choice of boundary conditions to approximately reproduce the KS-throat everywhere except close to the tip, where back-reaction effects deform the solution significantly at large field excursion.\par
The differential equations that follow from our ansatz are given in Appendix \ref{Appendix:EOM }. With the boundary conditions \eqref{pformboundaryconditions}, \eqref{metricboundaryconditions}, \eqref{fieldexcursionboundarycondition} and \eqref{axio-dilaton_fieldexcursion} a full numerical solution can be obtained. With the prescription to read off the effective $4d$ potential \eqref{4dbackreactedpotential} it can then be determined if our model can be used for realistic inflation.\par
Finally let us note that we have parametrized our ansatz such that at $\psi=\mathcal{O}(1)$ back-reaction effects become strong. The $C_2$-axion $\phi_{UV}\equiv \frac{1}{2\pi\alpha'}\int_{S^2} \delta C_2(\tau_{UV})$ with natural periodicity $\phi_{UV}\longrightarrow \phi_{UV}+2\pi$ can be expressed in terms of the field $\psi$ by $\phi_{UV}=2M\psi$ and the axion-decay constant\footnote{In our convention the periodicity of the canonically normalized axion is $\phi_c\longrightarrow \phi_c+2\pi f$.} $f$ can be estimated to be
\begin{equation}
\label{axiondecayconst}
\frac{f^2}{M_{pl}^2}\approx 2\pi^2(g_s \alpha')^2{\omega_2}_{MN}{\omega_2}^{MN}{\Big|}_{\tau_{UV}}\approx \frac{1}{32 M
^2}\left(B^{-4}{\Big|}_{\tau_{UV}}+C^{-4}{\Big|}_{\tau_{UV}}\right)\overset{\tau_{UV} \gg 1}{\approx} \frac{4}{3M^2}\frac{2^{1/3}}{\tau_{UV}-1/4}\, ,
\end{equation}
where $\omega_2$ is the harmonic $2$-form of $T^{1,1}$ normalized to $\int_{S^2}\omega_2=1$ and $B$ and $C$ are dimensionless radial functions that appear in the KS metric (see \eqref{zehnbein}). Here we have assumed that the $2$-cycle is of constant size as one passes from one throat to the other through the compact CY. The canonical field excursion in $4d$ can then be related to the variable $\psi$ by $\phi_c=f\phi_{UV}=(2Mf)\psi$ and $(2Mf)^2\approx \frac{16}{3}\frac{2^{1/3}}{\tau_{UV}-1/4}M_{pl}^2$. To get an upper bound on the throat length that is needed to generated the desired hierarchy let us assume that warping is the only effect that lowers the mass scale of the ultra-light mode\footnote{By this we mean that the Supergravity approximation is marginally valid at the tip of the throat, $g_sM\sim\mathcal{O}(1)$, and the compact CY is of the same size as the UV-region of the throat.}. Then in order to achieve $m^2/M_{pl}^2\approx 10^{-10}$, $\tau_{UV}\overset{!}{\approx} 20$ which implies $2Mf\approx 0.58M_{pl}$. Hence, back-reaction becomes strong at field excursion $\phi_c\approx 0.58 M_{pl}$. At this field excursion, the axion has already traversed $\frac{M}{\pi}$ natural periods. Because back-reaction effects are weak when $\psi\ll 1$ the inflaton can go through many of its periods with full computational control if $M$ is suitably large\footnote{Note that the Supergravity approximation becomes better at large $M$ because the typical length scale at the tip is $R^2\sim g_sM \alpha'$ such that the Supergravity approximation is good when both $g_s M\gg 1$ and $g_s\ll 1.$}.\par 
\subsection{Comparison with the Wilson-Contour of $B_2$}
We have argued that the integral of $C_2$ over the $2$-cycle of $T^{1,1}$ gives rise to an ultra-light mode. Moreover we claim that all other modes are stabilized at least at the warped KK-scale $m_{wKK}$. For this we assume of course that the axio-dilaton is stabilized in the UV by ISD-fluxes on cycles not relevant to the local KS-throat geometry. Perhaps the most obvious candidate that naively seems to be similarly light is the analogous ansatz for $B_2$. This however does not give rise to a similarly light mode as can be seen from \eqref{SUGRAeomC4}. Integrating this over the throat from the tip up to a position $\tau_1$ (let us call this region $\mathcal{C}_{\tau_1}$) yields:
\begin{equation}
\label{totalD3charge}
\int\limits_{\mathcal{C}_{\tau_1}}H_3\wedge F_3=\int\limits_{\mathcal{C}_{\tau_1}}dF_5=\int\limits_{T^{1,1}|_{ \tau=\tau_1}}F_5\equiv (2\pi)^4\alpha'^2 N(\tau_1)=16\pi^3g_s(M\alpha')^2 l(\tau_1)\, ,
\end{equation}
where $N(\tau_1)$ is the D3-brane charge integrated over $\mathcal{C}_{\tau_1}$ and we have used that $\del \mathcal{C}_{\tau_1}=T^{1,1}|_{ \tau=\tau_1}$ in the second step. Inserting our ansatz (far from the tip) $\delta C_2\sim f(\tau)\omega_2$ (with harmonic two form $\omega_2$) in \eqref{totalD3charge}, the LHS is left unchanged because $F_{3,KS}\longrightarrow F_{3,KS}+d\delta C_2$ and $d\delta C_2\sim f'(\tau)d\tau \wedge \omega_2$ wedges to zero with $H_3$ of the KS/KT-solution. It therefore does not enter as a source for additional $5$-form flux in \eqref{SUGRAeomC4}. This can also be seen from \eqref{lC2} which is left unperturbed far from the tip where $f\approx g$.\par 
Choosing the same ansatz for $\delta B_2$ instead, there is in contrast a non-trivial wedge-product between $d\delta B_2$ and $F_3$ of the KS-solution such that
\begin{equation}
N(\tau)-N^{KS}(\tau)\sim  \int\limits_{0}^{\tau}d\tilde{\tau}f'(\tilde{\tau})=f(\tau)\, .
\end{equation}
This can again be directly seen in \eqref{SUGRAeomC4}: as $(j,k)\longrightarrow (j+f,k+f)$, $l\longrightarrow l+f$.\par
The five-form flux thus changes linearly in the perturbation also far away from the tip of the throat. It then enters as a source on the RHS of \eqref{SUGRAeomB2}, 
\begin{equation}
F_5\wedge F_3\longrightarrow F_5\wedge F_3+\delta (F_5\wedge F_3)\, ,\quad \delta (F_5\wedge F_3)\propto f(\tau)
\end{equation}
which takes significant (i.e. not exponentially suppressed) values in the whole bulk. This can be interpreted as an effective $5d$ mass term for the $B_2$ axion and therefore leads to a $4D$-mass of order of the warped KK-scale $m_{wKK}$.
\section{Conclusion}
\label{Conclusion}
In this paper we presented a new idea for axion monodromy inflation in which the inflaton is the lightest Kaluza-Klein mode of the RR-2-form potential $C_2$ wrapped on a homologically trivial 2-cycle. One of the crucial technical points is that the mass of the lightest Kaluza-Klein mode is exponentially lower than that of the next excited mode, thus making this mode an ideal inflaton candidate. The monodromy arises due to the homological triviality of the 2-cycle similar to models proposed in \cite{F-term-inflation}, rather than due to a coupling to branes. Consequently, our construction does not require the presence of brane-antibrane pairs, thus avoiding the associated back-reaction issues \cite{Flauger:2009ab,BackreactionConlon}. Crucially, the exponential mass-suppression is due to the $S^2$ shrinking to zero size only in IR regions. This is why we base our model on the `double throat' shown in Figure \ref{fig:DoubleThroat}. Because back-reaction on other Supergravity fields cannot be neglected at large field excursion the non-linear Supergravity equations that govern the back-reaction are derived. Their numerical evaluation is left for future research.

The full type IIB Supergravity equations also show that the shift-symmetry of $C_2$ is preserved in the warped background except for the small monodromic potential that is generated at the tip. This in contrast is not true for the analogous Wilson-contour of $B_2$.

The perhaps most obvious open question that needs to be addressed by future research is the numerical solution of the equations of motion that were derived. From this the back-reacted potential and the maximal controlled field excursion of the model can be inferred. Naturally then, the question arises what the KS-throat decays to once the field excursion is set beyond its critical value.

Cornering the question of back-reaction from various dual descriptions seems to be a promising path towards gaining analytical insight: In the dual gauge theory the Wilson contour of $C_2$ at fixed radial position corresponds to the difference of $\theta$-angles of the two gauge group factors. Since $\theta$-angles are left unchanged under the cascade of Seiberg-dualities \cite{KlebanovStrassler}, the nearly constant profile of $C_2$ seems indeed to correspond to the correct Supergravity dual. It would be interesting to gain analytical insight into the back-reaction of our mode through this dual picture. Furthermore, in a T-dual type IIA picture the Wilson contour of $B_2$ corresponds to the distance between two $\text{NS}5$ branes with $D4$-branes suspended between them \cite{GravitywavesLinearInflation}. An analogous interpretation for the Wilson contour of $C_2$ that makes the monodromy manifest would be desirable and could also help address the question of back-reaction analytically.

It remains to be seen if the back-reacted potential extracted from the numerical analysis of the type IIB equations of motion is compatible with inflation in general and current observational constraints in particular. A simple quadratic potential is strongly disfavored by the latest data \cite{TensorToScalar}. Back-reaction effects however generically lead to a flattening of the potential \cite{10114521, 14053652} such that the model may well be in accord with current data.

Overall, we observe that our proposal realizes axion monodromy for a fairly minimal amount of ingredients. Given this relative simplicity and the high level of sophistication with which throat geometries can be controlled \cite{Frey:WarpedSpectroscopy,Frey:DimensionalReduction}, we expect our model to be a promising arena for further investigations into the viability of large field inflation in string theory. Regardless of the phenomenological implications, we would even like to hope that the possibility of large field inflation could be firmly established based on our simple scenario.\\
\\
\textbf{Acknowledgments}

We would like to thank Andrew Frey, Eran Palti, Fabrizio Rompineve, Patrick Mangat and Matthew Kleban for insightful discussions. This work was partly supported by
the DFG Transregional Collaborative Research Centre TRR 33 ``The Dark Universe''. LW is very grateful to Birzeit University for kind hospitality while this paper was finished. 
JM also acknowledges support from the Studienstiftung des deutschen Volkes. The work of AW is supported by the ERC Consolidator Grant STRINGFLATION under the HORIZON 2020 contract no. 647995.
\newpage
\appendix
\section{The Equations of Motion}
\label{Appendix:EOM }
In terms of the functions that parameterize our ansatz, we define
\begin{equation}
\begin{split}
*F_3^{(1)}=&\frac{B^2}{C^2}\left(g' -E(1-F)\right)
\, , \quad *F_3^{(2)}=\frac{C^2}{B^2}\left(f'-EF\right)
\, ,
\end{split}
\end{equation}
\begin{equation}
\begin{split}
*F_3^{(3)}=&-F'+E\frac{g-f}{2}
\, ,\quad *F_3^{(4)}=\frac{B^2}{C^2}\left(-\left(1+E^2\right)(1-F)+Eg'\right)
\, ,
\end{split}
\end{equation}
\begin{equation}
\begin{split}
*F_3^{(5)}=&\frac{C^2}{B^2}\left(-\left(1+E^2\right)F+Ef'\right)
\, ,\quad *F_3^{(6)}= \left(1+E^2\right)\frac{g-f}{2}-EF'
\, ,
\end{split}
\end{equation}
and furthermore define $*H_3^{(i)}$ which are obtained from $*F_3^{(i)}$ by the (symbolic) replacement
\begin{equation}
\label{replacement}
\begin{split}
F&\rightarrow -b\, ,\quad 
(1-F)\rightarrow b\, ,\quad
(g-f)\rightarrow (k-j)\, ,\\
f'&\rightarrow j'\, ,\quad
k'\rightarrow k'\, ,\quad
F'\rightarrow -b'\, .
\end{split}
\end{equation}
Let us further define the quantities
\begin{equation}
\label{gen.deriv}
\begin{split}
\{Z_g,Z_f, -Z_F,-X_{1-F},-X_{F},X_{g-f}\}^i&=* F_3^{(i)}-c* H_3^{(i)}\, ,\\
\{Z_k,Z_j,Z_b,X_b,X_{-b},X_{k-j}\}^i&=e^{-\phi}*H_3^{(i)}\, .
\end{split}
\end{equation}
\eqref{SUGRAeomC2} results in the three equations
\begin{equation}
\label{F3eom}
\begin{split}
&\left\{\frac{(A^4 Z_g)'}{A^4}-X_{g-f},\frac{(A^4 Z_f)'}{A^4}+X_{g-f},\frac{(A^4 Z_F)'}{A^4}-\frac{1}{2}(X_{1-F}-X_F)\right\}=- \frac{l}{4B^2C^2}\left\{-b,\quad b,\quad\frac{k-j}{2}\right\}\, ,
\end{split}
\end{equation}
while \eqref{SUGRAeomB2} leads to
\begin{equation}
\label{H3eom}
\begin{split}
&\left\{\frac{(A^4 Z_k)'}{A^4}-X_{k-j},\frac{(A^4 Z_j)'}{A^4}+X_{k-j},-\frac{(A^4 Z_b)'}{A^4}+\frac{1}{2}(X_{b}-X_{-b})\right\}= \\
&c'\left\{Z_g,\quad Z_f,\quad-Z_F\right\}+\frac{l}{4B^2C^2}\left\{F+bc,\quad 1-F-bc,\quad\frac{g-f}{2}-c\frac{k-j}{2}\right\}\, .
\end{split}
\end{equation}
Next, we define
\begin{equation}
\begin{split}
{F_3}_{(A)}=
&\{\frac{C}{B}F,\quad \frac{B}{C}(1-F),\quad\frac{g-f}{2},\quad  \frac{C}{B}\left(f'-EF\right),\\
&\frac{B}{C}\left(g'-E(1-F)\right),\quad F'-E\frac{g-f}{2}\}_A\, ,
\end{split}
\end{equation}
with $A=1,...,6$.
Furthermore ${H_3}_{(A)}$ are again obtained by the replacement \eqref{replacement}. Analogously to \eqref{gen.deriv}, we define the quantities
\begin{equation}
\label{gen.deriv2}
\begin{split}
\{P_{F},P_{1-F}, P_{g-f},Q_{f},Q_{g},Q_{F}\}_A&= {F_3}_{(A)}-c\cdot{H_3}_{(A)}\, ,\\
\{P_{-b},P_b,P_{k-j},Q_j,Q_{k},Q_{-b}\}_A&=e^{-\phi}{H_3}_{(A)}\, .
\end{split}
\end{equation}
Beware that $-P_b\neq P_{-b}$!
Then, \eqref{SUGRAeomC0} results in
\begin{equation}
\label{c0eom}
-4A^{-4}(c' A^4B^2C^2)'=
P_FP_{-b}+P_{1-F}P_b+2P_{g-f}P_{k-j}+Q_{f}Q_j+Q_{g}Q_{k}+2Q_{F}Q_{-b}\, ,
\end{equation}
and \eqref{SUGRAeomPhi} implies
\begin{equation}
\label{dilatoneom}
\begin{split}
&-32A^{-4}((e^{-2\phi})' A^4B^2C^2)'+8B^2C^2\left(c'\right
)^2=\\
&P_{-b}^2+P_{b}^2+2P_{k-j}^2+Q_j^2+{Q_{k}}^2+2{Q_{-b}}^2-P_F^2-P_{1-F}^2-2P_{g-f}^2-{Q_{f}}^2-{Q_{g}}^2-2{Q_{F}}^2\, .
\end{split}
\end{equation}
Thus, \eqref{F3eom} and \eqref{H3eom} determine the functions $f,g,F$ and $j,k,b$, while \eqref{dilatoneom} and \eqref{c0eom} determine $\phi$ and $c$. The only missing ingredients are Einstein's equations. In zehnbein basis they read
\begin{equation}
\label{Einstein_equ2}
\begin{split}
R^a_b=\left(\frac{1}{2}\hat{H}_3^{(2)}+\frac{1}{2}g_s^2e^{2\phi}\hat{\tilde{F}}_3^{(2)}+\frac{1}{4}g_s^2e^{2\phi}\hat{F}_5^{(2)}-4\hat{d\phi}^{(2)}+\frac{1}{2}g_s^2 e^{2\phi}\hat{F}_1^{(2)}\right)^a_b-\frac{1}{8}\delta^a_b(|H_3|^2+g_s^2e^{2\phi}|\tilde{F}_3|^2)\, ,
\end{split}
\end{equation}
where the terms on the RHS of \eqref{Einstein_equ2} take the following form
\begin{equation}
\begin{split}
e^{2\phi}g_s^2(\hat{\tilde{F}}_3^{(2)})^a_b&=(g_sM\alpha')^{-1}e^{2\phi}\left(\frac{1}{2DBC}\right)^2\text{diag}\left(0,0,0,0,F_{\tau}^{\tau}(\tau),F_{5}^5(\tau),F_{1}^1(\tau),...,F_{4}^4(\tau)\right)^a_b\\
&+(\delta^a_{\tau}\delta^5_b+\delta^a_5\delta^{\tau}_b)F_{\tau }^5+(\delta^a_{1}\delta^4_b+\delta^a_4\delta^{1}_b-\delta^a_{2}\delta^3_b-\delta^a_3\delta^{2}_b)F_{1}^4\, ,\\
F_{\tau}^{\tau}(\tau)&=Q_f^2+Q_g^2+2Q_F^2\, ,\quad 
F_{1}^1(\tau)=P_F^2+P_{g-f}^2+Q_f^2+Q_{F}^2\, ,\\
F_{3}^3(\tau)&=P_{1-F}^2+P_{g-f}^2+Q_g^2+Q_F^2\, ,\quad
F_{5}^5(\tau)=P_F^2+P_{1-F}^2+2P_{g-f}^2\, ,\\
F_{\tau}^5(\tau)&=Q_fP_F+Q_gP_{1-F}+2P_{g-f}Q_F
\, ,\quad
F_1^4(\tau)=-(P_F+P_{1-F})P_{g-f}-(Q_g+Q_f)Q_F\, ,
\end{split}
\end{equation}
while the $H_3$-part is determined by \eqref{replacement}.\par
The $5$-form field strength contributes as
\begin{equation}
\begin{split}
g_s^2e^{2\phi}\left(\hat{F}_{5}^{2}\right)^a_b&=(g_sM\alpha')^{-1}e^{2\phi}\left(\frac{l}{4DB^2C^2}\right)^2\text{diag}(-1,-1,-1,-1,-1,1,1,1,1,1)^a_b\, ,
\end{split}
\end{equation}
and the axio-dilaton as
\begin{equation}
\begin{split}
g_s^2e^{2\phi}(\hat{F}_1^{(2)})^a_b&=(g_sM\alpha')^{-1}e^{2\phi}\delta^a_{\tau}\delta^{\tau}_b \left(\frac{c'}{D}\right)^2\, ,\quad 
(\hat{d\phi}_1^{(2)})^a_b=(g_sM\alpha')^{-1}\delta^a_{\tau}\delta^{\tau}_b \left(\frac{\phi'}{D}\right)^2\, .
\end{split}
\end{equation}
Furthermore
\begin{equation}
\begin{split}
|H_3|^2+g_s^2e^{2\phi}|\tilde{F}_3|^2=&(g_sM\alpha')^{-1}\left(\frac{1}{2DBC}\right)^2e^{2\phi}[...]\, ,\\
\text{where}\,\, [...]=P_{-b}^2+P_b^2+&2P_{g-f}^2+Q_j^2+Q_k^2+2Q_{-b}^2+P_F^2+P_{1-F}^2+2P_{g-f}^2+Q_f^2+Q_g^2+2Q_F^2\, ,
\end{split}
\end{equation}
while the Ricci-tensor can be computed to give
\begin{equation}
\begin{split}
R^a_b&=(g_sM\alpha')^{-1}\text{diag}\left(R_{4D}(\tau),...,R_{4D}(\tau),R_{\tau}^{\tau}(\tau),R_{5}^5(\tau),R_{1}^1(\tau),...,R_{4}^4(\tau)\right)^a_b\\
&+(\delta^a_{\tau}\delta^5_b+\delta^a_5\delta^{\tau}_b)R_{\tau }^5+(\delta^a_{1}\delta^4_b+\delta^a_4\delta^{1}_b-\delta^a_{2}\delta^3_b-\delta^a_3\delta^{2}_b)R_{1}^4\, ,
\end{split}
\end{equation}
where
\begin{equation}
\begin{split}
R_{4D}(\tau)&=-\frac{1}{4A^4B^2C^2D^2}((A^4)'B^2C^2)'\, ,
\end{split}
\end{equation}
\begin{equation}
\begin{split}
R_{\tau}^{\tau}(\tau)=&-4\frac{1}{AD}\left(\frac{A'}{D}\right)'-2\frac{1}{DB}\left(\frac{B'}{D}\right)'-2\frac{1}{DC}\left(\frac{C'}{D}\right)'-\frac{1}{D^2}\left(\frac{D'}{D}\right)'-\frac{E^2}{4D^2}\left(\frac{B}{C}-\frac{C}{B}\right)^2
\, ,
\end{split}
\end{equation}
\begin{equation}
\begin{split}
R_{1}^1(\tau)=R_2^2(\tau)=&\frac{1}{B^2}-\frac{1}{2}\frac{D^2}{B^2C^2}+\frac{1+E^2}{8D^2}\left(\frac{B^2}{C^2}-\frac{C^2}{B^2}\right)-\frac{1}{2A^4B^2C^2D^2}\left[A^4C^2(B^2)'\right]'\, ,
\end{split}
\end{equation}
\begin{equation}
\begin{split}
R_{3}^3(\tau)=R_{4}^4(\tau)=&\frac{1}{C^2}-\frac{1}{2}\frac{D^2}{B^2C^2}+\frac{1+E^2}{8D^2}\left(\frac{C^2}{B^2}-\frac{B^2}{C^2}\right)-\frac{1}{2A^4B^2C^2D^2}\left[A^4B^2(C^2)'\right]'\, ,
\end{split}
\end{equation}
\begin{equation}
\begin{split}
R_{5}^5(\tau)=&\frac{1}{2D^2}+\frac{D^2}{B^2C^2}-\frac{1}{4D^2}\left(\frac{C^2}{B^2}+\frac{B^2}{C^2}\right)-\frac{1}{A^4B^2C^2D^2}\left(A^4B^2C^2\frac{D'}{D}\right)'\, ,
\end{split}
\end{equation}
\begin{equation}
\begin{split}
R_{\tau}^5=&\frac{E}{4D^2}\left(\frac{B^2}{C^2}+\frac{C^2}{B^2}-2\right)\, ,
\end{split}
\end{equation}
and
\begin{equation}
\begin{split}
R_1^4=&\frac{1}{4A^4B^2C^2D^2}\left[\frac{C}{B}\left(A^4B^4E\right)'-\frac{B}{C}\left(A^4C^4E\right)'\right]
\, .
\end{split}
\end{equation}
There are now seven Einstein equations for the five functions $A,B,C,D,E$. If we had allowed for a term $d(\tau)(g^1g^4-g^2g^3)$ in the $10d$ metric and $\Theta^{\tau}\propto L(\tau)d\tau$ with $L(\tau)\neq D(\tau)$ there would be seven equations for seven functions. Because such a more general ansatz can always be brought to the form that we have specified by a suitable coordinate re-parametrization the seven Einstein equations are not all independent and we do not expect the resulting system of differential equations to be overdetermined.
\newpage
\bibliography{axionKKrv}
\bibliographystyle{JHEP}
\end{document}